\documentclass[12pt,twoside]{article}
\usepackage[T1]{fontenc}
\usepackage[latin1]{inputenc}
\usepackage{times}
\usepackage{graphicx}
\usepackage{a4wide}

\title{Design of the Pluto Event Generator\footnote{Presented at the 17th International Conference on Computing in High Energy and Nuclear Physics}}
\DeclareOldFontCommand{\tt}{\normalfont\ttfamily}{\mathtt}

\author{I. Fr\"ohlich$^1$, T.~Galatyuk$^1$, R.~Holzmann$^2$, J.~Markert$^{1,2}$, \\
  B.~Ramstein$^3$,  P.~Salabura$^{2,4}$ and J.~Stroth$^{1,2}$}

\date{}
\begin{document}

\maketitle

$^1$ Institut f\"{u}r Kernphysik, Goethe-Universit\"{a}t, 60438 ~Frankfurt,
  Germany \\
$^2$ GSI Helmholtzzentrum f\"ur Schwerionenforschung GmbH, 64291~Darmstadt, Germany \\
$^3$ Institut de Physique Nucl\'{e}aire d'Orsay, CNRS/IN2P3, 91406~Orsay Cedex,
  France \\
$^4$ Smoluchowski Institute of Physics, 
  Jagiellonian University of Cracow, 30-059~Krak\'{o}w, Poland

\begin{abstract}

We present the design of the simulation package Pluto, aimed at the
study of hadronic interactions at SIS and FAIR energies.  Its main
mission is to offer a modular framework with an object-oriented
structure, thereby making additions such as new particles, decays of
resonances, new models up to modules for entire changes easily
applicable. Overall consistency is ensured by a plugin- and
distribution manager.  Particular features are the support of a
modular structure for physics process descriptions, and the
possibility to access the particle stream for on-line modifications.
Additional configuration and self-made classes can be attached by the
user without re-compiling the package, which makes Pluto extremely
configurable.

\end{abstract}

\section{Introduction}

Due to the fact, that experimental setups are usually not suited to
cover the complete complete phase space, event generators are very
important tools for experiments.  Practically, the experimental
physicist needs a tool in hand which allows to control almost all
kinematic variables with a manageable user interface and to exchange
the physics models which have to be compared to measured data.

In our contribution, we present the software structure of the Pluto
event generator~\cite{pluto} originally developed for the HADES
experiment~\cite{nim} but successfully used by other collaborations in
the hadronic physics field as well.  Its recent redesign partly
discussed in this contribution enhanced its flexibility and provide
new features which allow to meet new challenges coming with the
detector studies for the new FAIR experiments PANDA~\cite{panda} and CBM~\cite{cbm}.

Pluto is a collection of C++ classes, adding up to the framework of a
simulation package for hadronic physics interactions in the energy 
regime up to a few GeV.  It is launched interactively within the
ROOT~\cite{root} environment, and makes use of ROOT only, without
requiring additional packages. The focus is on streamlining particle
generators by providing the tools to set up and manipulate particles,
reaction channels, and complex reactions, as well as applying
experimental filters on the reaction products, such as geometrical
acceptance and kinematical conditions. Typical simulations may be
executed with a few lines of input, with no expertise required on the
part of the user.  It standardizes simulations by providing a common
platform that is accessible via the analysis environment (ROOT), with
a straightforward user interface that does not inhibit non-experts in
simulations.

The output may be analyzed on line, or further forwarded to
a digitization package. The Pluto framework includes models for hadronic and electromagnetic
decays, resonance spectral functions with mass-dependent
widths, and anisotropic angular distributions for selected channels. A
decay-manager interface enables "cocktail" calculations. An extensive
particle data base is available, with capabilities to support
user-defined ones. Various particle properties and decay modes are
included in the data base. Thermal distributions are implemented,
enabling multi-hadron decays of hot fireballs.

The general philosophy of Pluto is that on one hand the users should
be able to drive the package in a very easy way, but on the other hand
are allowed to customize it without much restrictions. This means that
the steering application has the opportunity to add new decays/particles among with new
models up to the incorporation of plugins\footnote{The PC gaming industry, 
  making benefit from charge-free down-loadable packages, 
  calls such extension features also ``mod''.}.

\begin{figure}[h]
  \begin{center}
    \includegraphics[width=0.8\columnwidth]{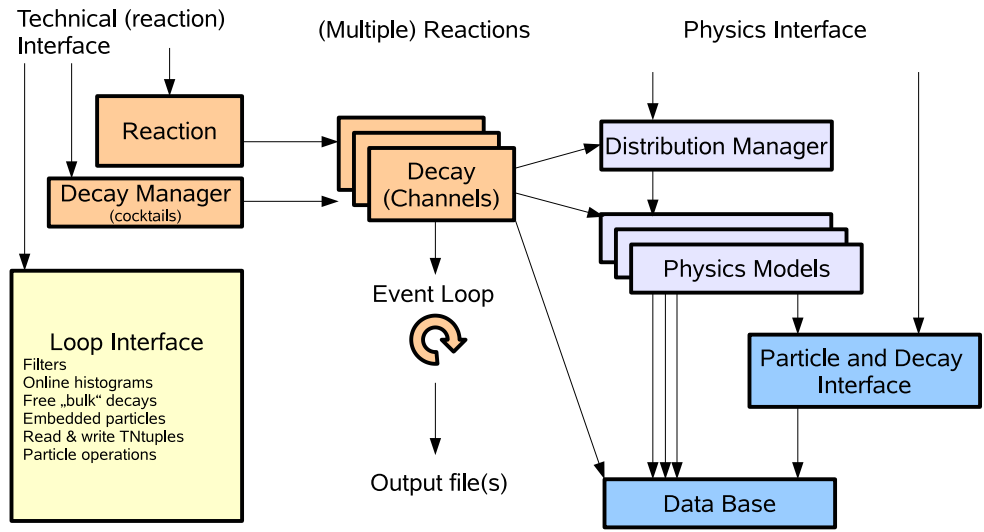}
    \caption{\label{fig:pluto_at_a_glance} Structure and design of
      Pluto: the (orange) boxes in the upper left show the user classes to set up and
      handle reactions (and decay channels). 
      The (yellow) box on the left is the interface for accessing the
      particle stream while the event loop is running. 
      The (blue shaded) boxes in the lower right corner represent the data
      base for the particles and decays among with a user interface
      for customization. The (light blue) boxes above are the distribution
      interface and the included physics models (color online).}
  \end{center}
\end{figure}

The structure and design of Pluto are sketched in
Fig.~\ref{fig:pluto_at_a_glance}.  The framework can be roughly
divided into 3 parts: a user interface for event production, a physics
package, and an event loop interface. In this report we would like to
concentrate on the latter two issues: the first one allows to add
and/or select physics cases, where models can be either build-in
models or runtime replacements, which will be discussed in
Sec.~\ref{sec:models}.  This area of Pluto works even outside the
event loop for the calculation of parameters (mass distributions,
decay widths,...). The other block allows for external particle
decay and acceptance filter classes (see Sec.~\ref{sec:bulk}) and
filling of on-line histograms (see Sec.~\ref{sec:projector}).

As the interface for event production is described in detail
elsewhere~\cite{pluto} in this context we will give only a short example.
After all models have been configured accordingly, an elementary 
reaction chain is defined
with a character string containing the reaction products. To perform, e.g., the
$\Delta^+$ production in the $pp$ reaction with a consecutive Dalitz decay as
discussed in the following sections, 
the commands as defined below are sufficient:
\begin{verbatim}
 PReaction *r = new PReaction("1.25","p","p",
                              "p D+ [p dilepton [e+ e-] ]"); 
 r->Print(); 
 r->Loop(10000);
\end{verbatim}
with the number 1.25 as the kinetic beam energy in GeV, the beam and
target particle indicated by following two options and finally a
string describing the decay.
Particles are defined via their data base name (as "{\tt p}" for a
proton or "{\tt D+}" for a $\Delta^+$).  The setup of the reaction channels
and the attachment of all selected distributions and models is done by
the framework, with the possibility
to dump the complete chain and its models via the {\tt Print()} method.
Finally, the command {\tt Loop(n)} produces $n$ events.

Then, event-event-by-event, the individual steps in each reaction are
executed via a common interface of all classes making distributions (see~\cite{pluto} for details)
which do the calculation of final particle tracks (also the unstable ones)
which are stored in an array.

\section{The models}\label{sec:models}

\subsection{Architecture}

In the Pluto framework, each physics process is described by a number
of objects, instantiated from dedicated classes. Beside the objects
which communicate with the event loop and set the 4-momenta of
the resulting particle tracks (to form the above-mentioned ``distributions''), Pluto offers 
methods to make calculations 
including various channels and their accompanied functions 
in advance to (or event without) the 
event-loop.  For this purpose, a relative data base has been included,
which connects particles and decays to ``models''. At least one
(so-called ``{\em primary}'') model is needed  per decay in order 
to describe the respective
mass shape (usually calculated by a mass-dependent Breit-Wigner in a
recursive way) and the partial/total decay width. Each selected model
can be accessed from all locations of the software package as well as
by the user.

The choice which of the allover offered physics models are linked to the data
base can be done inside the production ROOT-macro by calling a distribution
manager (for details about the handling see~\cite{pluto}), which itself
takes care that only one primary model per decay and
particle is selected, respectively.
Moreover,
Pluto offers the user the freedom to attach user-defined models based
on its own sources or even new classes (written in C++) at runtime,
thereby replacing built-in models. These features make Pluto an
extremely versatile and highly configurable tool for simulations,
while guaranteeing overall consistency at all times.

One specific feature of Pluto is that we do not use monolithic decay
models only but allow for the splitting of the underlying physics process into different models in a
very granular way (e.g., to exchange form factors or total cross
sections). This turned out to be a very important tool in order to
check various scenarios along with measured data. Therefore Pluto
allows for the attachment of secondary models for all kinds of
purposes. Here, a secondary model is an object for a particle/decay
returning a (complex) number as a function of a defined number of
values.  We do not attempt to have a base class for each kind of
function and model (e.g., a total cross section base class, a base
class for $d\Gamma / dm$, etc.)  because this makes further extensions
difficult.  Our approach is as follows: for each class of secondary
models a new entry (defined by a unique name) is  added
dynamically to the data base. Consequently, each model is bound to two data base keys: 
one to link it to a decay (or particle)  and
one more for the type of secondary model.

Anyhow, there the user does not need to take care about 
these issues, as the following template example shows.

\subsection{Template example}

In Pluto, each model is inherited from the base class {\tt PChannelModel}
with a lot of useful features. 
In order to offer the result of the calculation
the {\tt GetWeight()} method has to implemented. Let us create such
an example class as a template for the following discussion and future developments:
\begin{verbatim}
 class PHelperFunctionTest : public PChannelModel {
 public:
   PHelperFunctionTest(const Char_t *id, const Char_t *de, 
                       Int_t key);  
   Double_t GetWeight(Double_t *x, Int_t *opt=NULL);
 }
\end{verbatim}
with an realization of the (in this case 1-dimensional) function:
\begin{verbatim}
 Double_t PHelperFunctionTest::GetWeight(Double_t *x, 
                                         Int_t *opt) {
   Double_t xx = x[0];
   val = ...
   return val;
 };
\end{verbatim}

Objects can be instantiated to work as a model for specific physics processes by the usage of a 
parser residing inside the constructor of the base class:
\begin{verbatim}
 PHelperFunctionTest *obj = new
   PHelperFunctionTest("unique_name@D+_to_p_dilepton/helper",
                       "Test",-1);
\end{verbatim}
Here, the {\tt 
unique\_name} is used to identify the object inside the distribution 
manager.
The string hereafter binds the new object 
to a decay which is in this case $\Delta^+ \to p \gamma^*$. 
The optional appendix {\tt /helper} enables the new object to be a secondary model of this type.

Let us now assume, the primary model has to access its helper function. In this
case it is sufficient to execute the following command at a single place (an {\tt Init()}
function is available for that):
\begin{verbatim}
 PChannelModel *sec_model=GetSecondaryModel("helper");
\end{verbatim}
and execute the daughter method {\tt sec\_model->GetWeight(value)}
whenever required.  The access to alien secondary models is possible
as well but a further discussion on this topic would exceed the scope of this
report.

\subsection{Example: form factors}

The power of such a splitting can be demonstrated best with an real
example. For this purpose let us use the study of the electromagnetic $\Delta$ Dalitz
decay $\Delta \to N \gamma^* \to N e^+ e^-$. As explained in the introduction Pluto 
reads this as a
multi-step process: in the first step, the $\Delta$ mass is sampled using
a mass-dependent width. In the second step, the already sampled
$\Delta$ mass $m_{\Delta}$ is used by the following Dalitz decay model to sample the
di-electron (=virtual photon) mass  $m_{ee}$.

By using a dedicated plugin:
\begin{verbatim}
 makeDistributionManager()->Exec("dalitz_mod: krivoruchenko");
\end{verbatim}
the following model for the di-electron mass spectrum is enabled, 
using the prescription of
Ref.~\cite{wolf} with:

\begin{equation}\label{eqn:wolf}
\frac{d\Gamma^{\rm \Delta\to Ne^+e^-}_{m_\Delta}(m_{ee})}{dm_{ee}}
=\frac{2\alpha}{3\pi m_{ee}}
\Gamma^{\Delta \to N \gamma^*}_{m_\Delta}(m_{ee})
\end{equation}
and the decay rate given by Krivoruchenko {\it et~al}~\cite{kriv}:
\begin{eqnarray}\label{eqn:delta_dalitz_decay_kriv}
  \nonumber \Gamma^{\Delta \to N\gamma^*}_{m_\Delta}(m_{\gamma^*})
  & = &  \left( G^{\Delta \to N\gamma^*}_{m_\Delta}(m_{\gamma^*}) \right)^2 \\
  & & \nonumber \times \frac{\alpha}{16}
\frac{(m_\Delta + m_N)^2}{m_\Delta^3m_N^2}
\sqrt{y_+ y_-^3}, \\ y_\pm & =& (m_\Delta \pm m_N)^2-m_{ee}^2
\end{eqnarray}
where the index $N$ refers to the produced nucleon, $e$ is the electron
charge, and $G^{\Delta \to N\gamma^*}_{m_\Delta}(m_{\gamma^*})$ represents the
electromagnetic transition amplitudes. 
Here, from the software point of view, it is important to know that this factor
is a separated object in Pluto making the addition of new complementary descriptions very easy. This avoids
the usage of nasty flags which is error-prone. E.g.,
following the QED approach (assuming that no hadronic transition form factor is involved) from
Ref.\cite{kriv} gives:

\begin{equation}\label{eqn:form_factor}
  \left(G^{\Delta \to N\gamma^*}_{m_\Delta}(m_{\gamma^*})\right)^2
  = G_M^2 + 3G_E^2 + \frac{m^2_{\gamma^*}}{2m_\Delta^2} G_C^2
\end{equation}
with $G_M$, $G_E$, and $G_C$ as the magnetic, electric and Coulomb transition
form factors. 

Alternatively, one can switch to the VMD-like form factor of Iachello
and Wan~\cite{iachello} which is already implemented and will be
discussed in detail in a further, separate publication.

After activating this plugin, both above-mentioned versions (VMD and QED) 
are offered, appearing as objects with their unique names
in the directory-like distribution manager structure. 
\begin{verbatim}
makeDistributionManager()->Print("vmd");                  
--------------------
PDistributionManager
--------------------
  root                Root group: 494 objects (of 504), 
                        subgroups: 9
     particle_models  Mass sampling of particles : 27 objects 
                        (of 29)
     decay_models     Phase space mass sampling & decay partial
                        widths: 219 objects (of 225)
     polar_angles     Polar angles in elementary particle 
                        production: 10 objects (of 10)
     genbod_models    Momentum sampling: 220 objects (of 220)
         ...
     vmd              VMD form factors: 0 objects (of 2)
( )  D+_vmd_ff        VMD ff for D+ -> p e+e-
 (X) D+_qed_ff        QED ff for D+ -> p e+e-
         ...
     qed              QED form factors: 2 objects (of 2)
\end{verbatim}

Possible conflicts between the QED and VMD models are indicated.
The user can now activate all models inside the directory via: 
\begin{verbatim}
 makeDistributionManager()->Enable("vmd");
\end{verbatim}

%Coming back to the consistency issue: the change of the partial decay width
%changes also the mass-dependent branching ratio, therefore already the $\Delta$ mass
%sampling has to take the chosen model into account.

\section{The event loop interface}\label{sec:bulk}

\begin{figure}[t]
  \begin{center}
    \includegraphics[width=0.5\columnwidth]{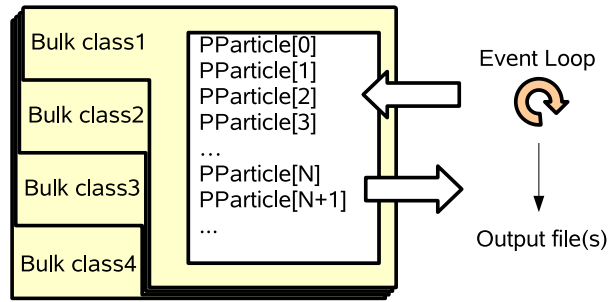}
    \caption{\label{fig:bulk_interface} The bulk interface.  }
  \end{center}
\end{figure}

The aim of the event loop interface is to allow access to the particle bulk
before it is written to disk by providing an interface for bulk decays (also using 3$^{rd}$ party
event generators such as Pythia~\cite{pythia}), file access, implantation of embedded 
particles for detector studies, and filter classes. 
It works - like indicated in Fig.~\ref{fig:bulk_interface} - by stacking
individual bulk classes (by {\tt PReaction::AddBulk(PBulkInterface * obj)}) 
and making use of the base-class {\tt Modify()} method which handles
the embedding and removing of particles or mark them as decayed/invalid.

Like in the previous section, 
user classes can be compiled - if needed - and added on-the-fly, which
could be very helpful in the context of FAIR experiments studies,
as the following applications show:

\subsection{File access:} Particle tracks might be added from 
 (or written to) user files in any self-defined format. This
feature makes Pluto open for other collaborations and 
enables it to be used as an ``afterburner'' for externally
created simulation data or output of transport models.

\subsection{Embedded particles:}
Test particles can be embedded into a background reaction which is an
important tool for detectors studies, e.g. to test tracking in a high
multiplicity environment.  Moreover, among with the previous feature
the mixing of embedded particles with either real events or
3$^{rd}$ party simulation is one of the applications. 
In the first case, the reconstructed vertex
of the analyzed experimental data is used to simulate rare probes (as the di-electrons) 
at exactly
the same vertex.  In this way the efficiency of these probes can be
calculated with the best background one can have, namely measured data.

\subsection{Bulk decay:}
In addition to a well defined reaction chain, Pluto offers free
``bulk'' decays, i.e.\ the decay of (embedded) particles following the
calculated mass-dependent branching ratios. At this stage it is possible as well to
employ 3$^{rd}$ party event generators. Only to give an example: Pluto can generate embedded
particles like resonances with a selected distribution and the Pythia
package (or Pluto itself) can be used for their realistic free decay.

\subsection{Detector acceptance class:} \label{sec:filters}
The particles created in the previous steps 
can be pre-filtered using a ``detector acceptance'' user class
following a class definition as outlined below:
\begin{verbatim}
 class PMyDetector : public PBulkInterface {
 private:
   Double_t *acc;
 public:
   Bool_t Modify(PParticle ** stack, int *decay_done, int * num, 
                 int stacksize);  
 };

 PMyDetector::PMyDetector() {
   acc = makeStaticData()->GetBatchValue("acc_value");
 };
\end{verbatim}
with {\tt acc\_value} as a booked batch value (see below for more information)
and, e.g.:
\begin{verbatim}
 Bool_t PMyDetector::Modify(PParticle ** stack, ... ) {
 *acc=1.;
 for (int i=0; i< *num; i++) {
   PParticle * cur = stack[i];
   if ("check for cur") *acc=0.;
   }
 }
\end{verbatim}

\section{Projecting events}\label{sec:projector}

\subsection{General idea}

The usual way to analyze generated events (beside a full Monte-Carlo
production including digitization) is to open the simulation file with an
appropriate analysis macro, loop over the number of events, read the
particle objects and do the required operations inside the
loop. E.g., in hadronic physics, in particular for exclusive studies,
usually missing and invariant masses are reconstructed, and possibly
after boost operations angular correlations are investigated.
In addition, 
signal selection has to be done not only on a single track level but using 
particle correlations (such as a invariant/missing mass selection).
 This
prevents to use the features provided by {\tt TTree::Draw()} only, as
these are fixed to single track properties. 

\begin{figure}[t]
  \begin{center}
    \includegraphics[width=0.6\columnwidth]{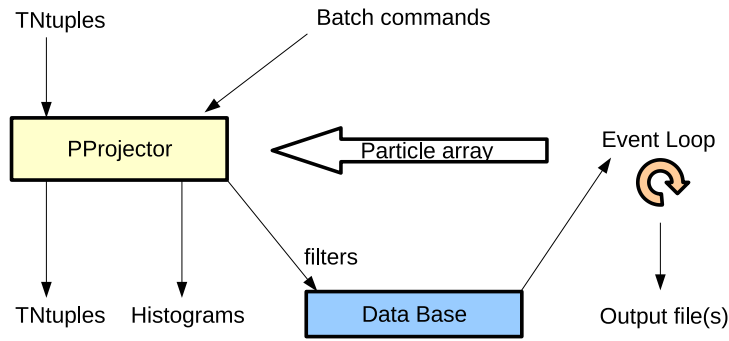}
    \caption{\label{fig:pluto_pprojector} Overview of the {\tt PProjector} design.  }
  \end{center}
\end{figure}

The main aim of the ``{\tt PProjector}'' (see Fig.~\ref{fig:pluto_pprojector})
 is to offer a simplified analysis method following a similar
idea as {\tt TTree::Draw()} in order to project particles (and their
correlations) onto on-line histograms and/or {\tt TNtuple}s by making
use of the bulk interface described in the previous section.  This
means, no analysis macro is needed for a large number of typical observables.

There are two ways to include the projector into the event loop. Either
one or more of its instances are added to the bulk interface (for an alternating 
mixing with other classes), or,
which is sufficient for the most cases, the short version {\tt PReaction::Do(char * command)} 
can be used, where {\tt command} follows a
C++-like (but more simple) syntax. Each of these commands - just forming steps in a batch -
is executed inside the event loop.

\subsection{Histograms}

To fill histograms one can use the extended method {\tt PReaction::Do(THx * histo, char *
  command)} where {\tt histo} points to a 1/2/3-dimensional histogram.

In this case the command has to define  the key variables  {\tt \_x,  \_y} and  {\tt \_x}
as the respective axis value, depending on the chosen dimension of
{\tt histo}. The Pluto particle array objects can be accessed via
its Pluto particle name inside brackets as {\tt [name,num]} with the optional {\tt num}
as the consecutive number of equal particles inside each event. All objects can be combined
with a number of build-in methods, moreover all browsable methods of the ROOT-class 
{\tt TLorentzVector} can be employed, as the following example indicates:
\begin{verbatim}
 r->Do(histo1,"_x = [D+]->M()");
\end{verbatim}
which fills the 1-dimensional  {\tt histo1} with the $\Delta^+$ mass.
Invariant masses (composite particles) can be formed 
by adding the objects:
\begin{verbatim}
 "_x = ([e+] + [e-])->M2()"
\end{verbatim}
and stored for a consequent use in semi-colon 
separated sub-commands without requiring a ``new'' operator: 
\begin{verbatim}
 "dilepton = ([e+] + [e-]); _x = dilepton->M2()"
\end{verbatim}

Let us now come to a more complex example, which is
the calculation of the polar emission angle of one of the outgoing protons
in a $p+p$ reaction (N.B. Pluto treats the c.m. system as a particle named  {\tt p + p}): 
\begin{verbatim}
 "p1 = [p,1]; p1->Boost([p + p]); _x = cos(p1->Theta())"
\end{verbatim}

In this example, in the first sub-command the first appearing proton is copied to 
the object {\tt p1} which is furthermore boosted into the c.m.\ energy frame and
in the last step the cosine of its polar angle is written as a value to the x-axis.

\subsection{TNtuple in- and output}

The projection  
to {\tt TNtuple}s turned out to be very useful as it 
allows to make additional selections (and definition of the histogram axis)
afterwards which is a very simple task using the {\tt  PProjector} 
following the syntax as described above. Let us now assume, the
momenta of simulated $\eta$'s should be stored. Then the first step would be to define the
 {\tt TNtuple} with branch names:
\begin{verbatim}
 TNtuple *ntuple = 
   new TNtuple("ntuple","...","eta_px:eta_py:eta_pz");
\end{verbatim}

Hereafter, the {\tt Do()}-method has to be called like in the above-indicated example 
but with a pointer to a {\tt TNtuple}-object instead of a histogram 
and the variable names of the ntuple instead of the axis names:
\begin{verbatim}
 "eta_px = [eta]->Px(); eta_py = [eta]->Py(); 
    eta_pz = [eta]->Pz()" 
\end{verbatim}

The filling of the ntuple is done by Pluto, only (but on purpose)
the opening and closing of the root-file has to be 
called externally.

In addition to the writing of {\tt TNtuple}s it is also possible to read (back) any
ntuple, modify its values and write it back or make histograms. 
Let us use the ntuple of the previous example which uses the ntuple (usually from a file) 
and defines an {\it empty} reaction, but attaches the ntuple as a 
3rd party input:
\begin{verbatim}
 PReaction my_reaction; 
 my_reaction.Input(ntuple);
\end{verbatim}

The next step is to use the branches of the ntuple in the batch command line to, e.g., reconstruct the
4-momentum:
\begin{verbatim}
 "myeta = P3M(eta_px,eta_py,eta_pz,0.54745)"
\end{verbatim}
with {\tt P3M} as the short syntax for a vector constructor  with 3 momenta and the particle mass as parameters.
At this point any kind of mathematical operation available via {\tt TFormula} can be applied as well.
Finally using {\tt Loop()} without a number reads all events from the ntuple.

\subsection{Filter commands}

The employment of batch commands allows to define on-line filters on
single particle properties as well as kinematic variables which
are accessible as indicated above. These filters can be used to select events for
single histograms only or forwarded to the event loop to apply
physical constraints, before a full event is written to disk, hence
reducing the event file. Variables prepared in a detector class as outlined in
Sec.~\ref{sec:filters} can be integrated very easily.

This is realized by using the {\tt if}-command, followed by the variable or
a combined condition based on particle observables. 
The remaining batch sub-commands are executed only under a positively 
evaluated condition.
It goes without saying that the condition statement can combine all methods among
with the established {\tt TFormula} syntax: 
\begin{verbatim}
 "if(obj->P() > 0.3 && obj->P() < 0.6); dowhatever..."
\end{verbatim}
In this case, histograms and ntuples
in the  {\tt Do()} method are only filled under the defined condition, or, ``dowhatever'' can 
be used to re-define variables.

In this context, it should be noted, that variables, starting
with ``{\tt \#}'' are interpreted by the Pluto event loop as a condition for the
final event file. 
 In our next case, an event is accepted if both the electron and
the positron are produced with polar angles in the range of 18 to 85
degrees: 
\begin{verbatim}
 "theta_ep = ([e+]->Theta() * 180.)/TMath::Pi()" 
 "theta_em = ([e-]->Theta() * 180.)/TMath::Pi()" 
 "#filter = 1; if(theta_ep<18 || theta_ep>85 || theta_em<18 
                  theta_em>85); #filter = 0"
\end{verbatim}
by this means it is only a single step to activate the 
detector class {\tt MyDetector} from Sec.~\ref{sec:filters}:
\begin{verbatim}
 "#acc_filter = acc_value"
\end{verbatim}
and optionally combine the prepared acceptance flag with our kinematical conditions.

\section{Summary} 

In summary, we presented new developments for the Pluto event generator
which moves this package from being a pure event generator to a comprehensive framework.
The particular features are the bookkeeping of various models and the user interface
for their simple selection. The new event loop interface allows to mix several sources, to define and enable
online selections, project to histograms and to make a lot of operations. Detector acceptance
and resolution can be included without much overhead, 
which is of particular importance for detector studies
in the context of the FAIR project.

In the near future, we will implement general purpose resolution scenarios using a common data format 
and 
file readers to transport models (like UrQMD~\cite{urqmd} and HSD~\cite{hsd}).
This will allow to test very rapidly the performance of different detector configuration options.

\section*{Acknowledgements}

This work was supported by the Hessian LOEWE initiative through the Helmholtz 
International Center for FAIR (HIC for FAIR). We would like to thank B. K\"ampfer for
reading and commenting the manuscript.


\begin{thebibliography}{9}

\bibitem{pluto}
I.~Fr\"ohlich {\it et al.},
  %``Pluto: A Monte Carlo Simulation Tool for Hadronic Physics,''
  PoS  {\bf ACAT2007}, 076 (2007)
  [arXiv:0708.2382 [nucl-ex]].
  %%CITATION = POSCI,ACAT2007,076;%%


\bibitem{nim}
  G.~Agakishiev {\it et al.}  [The HADES Collaboration],
  %``The High-Acceptance Dielectron Spectrometer HADES,''
  arXiv:0902.3478 [nucl-ex].
  %%CITATION = ARXIV:0902.3478;%%

\bibitem{panda}
  J.~G.~Messchendorp  [PANDA Collaboration],
  %``Hadron Physics with Anti-Protons: The PANDA Experiment at FAIR,''
  arXiv:0711.1598 [nucl-ex];
  %%CITATION = ECONF,C070910,123;%%
  B.~Seitz [PANDA Collaboration],
  %``The PANDA Project at FAIR,''
%\href{http://www.slac.stanford.edu/spires/find/hep/www?irn=8101868}{SPIRES entry}
{\it Prepared for 16th International Workshop on Deep Inelastic Scattering and Related Subjects (DIS 2008), London, England, 7-11 Apr 2008}.

\bibitem{cbm}
  V.~Friese,
  %``The CBM experiment at GSI/FAIR,''
  Nucl.\ Phys.\  A {\bf 774}, 377 (2006);
  %%CITATION = NUPHA,A774,377;%%
  P.~Senger, T.~Galatyuk, D.~Kresan, A.~Kiseleva and E.~Kryshen,
  %``Cbm At Fair,''
  PoS  {\bf CPOD2006}, 018 (2006);
  %%CITATION = POSCI,CPOD2006,018;%%
  V.~Friese,
  %``The CBM experiment at FAIR,''
  PoS  {\bf CPOD07}, 056 (2007).
  %%CITATION = POSCI,CPOD07,056;%%

  

\bibitem{root} http://root.cern.ch

\bibitem{wolf}
  G.~Wolf, G.~Batko, W.~Cassing, U.~Mosel, K.~Niita and M.~Schaefer,
  %``Dilepton production in heavy ion collisions,''
  Nucl.\ Phys.\  A {\bf 517} (1990) 615.
  %%CITATION = NUPHA,A517,615;%%

\bibitem{kriv} 
  M.~I.~Krivoruchenko and A.~Faessler,
  %``Comment on Delta radiative and Dalitz decays,''
  Phys.\ Rev.\  D {\bf 65}, 017502 (2002)
  [arXiv:nucl-th/0104045].
  %%CITATION = PHRVA,D65,017502;%%



\bibitem{iachello} 
  Q.~Wan and F.~Iachello,
  %``A unified description of baryon electromagnetic form factors,''
  Int.\ J.\ Mod.\ Phys.\  A {\bf 20} (2005) 1846;
  %%CITATION = IMPAE,A20,1846;%%
  Q. Wan, Ph.D. Thesis, Yale University, New Haven, Connecticut (2007);
  F. Iachello, priv. comm. (2008).

\bibitem{pythia}
  T. Sjostrand, S. Mrenna and  P. Skands,
  JHEP 0605 (2006) 026, [hep-ph/0603175].

\bibitem{urqmd}
M.~Bleicher {\it et al.},
  %``Relativistic hadron hadron collisions in the ultra-relativistic quantum
  %molecular dynamics model,''
  J.\ Phys.\ G {\bf 25}, 1859 (1999)
  [arXiv:hep-ph/9909407].
  %%CITATION = JPHGB,G25,1859;%%

\bibitem{hsd}
  W.~Cassing and E.~L.~Bratkovskaya,
  %``Hadronic and electromagnetic probes of hot and dense nuclear matter,''
  Phys.\ Rept.\  {\bf 308}, 65 (1999).
  %%CITATION = PRPLC,308,65;%%


\end{thebibliography}
\end{document}